# Proximity effect and electron transport in the oxide hybrid heterostructures with superconducting/magnetic interfaces

G.A. Ovsyannikov[1,2], K.Y. Constantinian[1], Yu.V. Kislinski[1], A.V. Shadrin[1], A.V. Zaitsev[1],
V.V. Demidov[1], I.V. Borisenko[1], A.V. Kalabukhov[2], D. Winkler[2]

[1] Kotel'nikov Institute of Radio Engineering and Electronics, Russian Academy of Sciences, 125009, Moscow, Russia.

[2] Department of Microtechnology and Nanoscience, Chalmers University of Technology, S41296, Gothenburg, Sweden

## Abstract

We report on the study of electron transport in the oxide heterostructures with superconductor/magnetic matter (S/M) interfaces where anomaly large penetration of superconducting correlations in magnetic matter (proximity effect) is realized. The developed theoretical model based on multilayer magnetic structure of M-interlayer and experiment show presence of the long-range proximity effect at S/M-interface with antiferromagnetic (AF) ordering of the interlayer magnetization. The investigated hybrid heterostructures include cuprate superconductor ($S_d$), the AF-cuprate interlayer and conventional superconductor (Nb). The superconducting critical current with density $1-10^3 A/cm^2$ and the characteristic voltage, $I_C R_N$ =100-200 µV ($I_C$ and $R_N$ are critical current and normal resistance, correspondingly) are observed at liquid helium temperature for 15-50 nm thick M-interlayer made of AF film $Ca_XSr_{1-X}CuO_2$. These heterostructures demonstrate deviation from sin-type superconducting current-phase relation and had the critical current of the second harmonic of 10-20% of the first harmonic one. The hybrid heterostructures with S/AF interface show high sensitivity to the external magnetic field that is possibly caused by the influence of an external magnetic field on the canting of magnetic moments of individual layers of AF -interlayer. Substituting the AF cuprate by a manganite film, no critical current was observed, although the M- interlayer was made very thin, down to 5 nm.

1. **Introduction**

Coexistence of superconducting and magnetic ordering in solids is of great interest for fundamental physical studies and electronic applications. The exchange mechanism of ferromagnetic order tends to align spins of superconducting pairs in the same direction preventing singlet superconducting pairing [1-2]. At the interfaces between superconducting (S) and magnetic (M) materials, however, the superconducting and magnetic correlations may interact due to the proximity effect (penetration of superconducting correlations into magnetic) resulting in interplay between superconducting and magnetic ordering and novel physical phenomena may appear. However, up to now the most of the activity were devoted to investigation of systems in which magnetic matter is a ferromagnetic (F). One of the important properties of the proximity effect in S/F interface is a damped oscillatory behavior of the condensate wave function induced in the F- layer. This may lead, in particular, to a π-phase shift [3] in the superconducting current-phase relation of S/F/S Josephson junctions experimentally demonstrated in [4]. In some structures, for instance, in F/S/F spin-valve junctions one can use F-layers to control superconductivity by the magnetic field or current due to changing of exchange field orientations in F-electrodes [5]. Much less attention was paid to investigation of superconducting structures with M-interlayer having AF ordering. Recently L. Gor'kov and V. Kresin [6] assumed a model of S/AF/S structure where an AF interlayer consists of F-layers with magnetizations aligned alternatively and placed perpendicular to the S-electrodes and the biasing current is directed along the layers (G-K model). The model predicts higher critical current than in S/F/S structure and show that even a minor canting of magnetic moments in the presence of magnetic field causes a noticeable oscillations of the critical current. The authors of the work [7] considered an AF interlayer as a series of F-layers placed in parallel to the S/AF interface and showed that the Josephson critical current significantly depends on whether the number of layers is odd or even.

Theoretical investigation of a S/M/S structure with M-interlayer composed of F-layers each one with a thickness significantly exceeding the atomic-scale (magnetic multilayer structure, MMS) was carried out in the work [8] The orientation of the F-layers in MMS-model was parallel to the S/M interface. It was shown that for AF- ordering of the magnetizations in the M-structure the significant dependence of the proximity effect will take place and amplitude of Josephson current depends on N number of layers.

Experimental observation of the Josephson effect in Nb/Cu/FeMn/Nb polycrystalline thin film multilayerstructures has been demonstrated in [9], where γ-$Fe_{50}Nb_{50}$ alloy is used as metallic AF interlayer. Significant suppression of superconductivity has been observed in S/AF bilayer with FeMn layer, the critical current modulation with magnetic field $I_c(H)$ of the S/AF/S structure shows rather conventional Fraunhofer pattern [9]. If, instead of polycrystalline metallic AF material one substitutes it by an array of ferromagnetic layers with alternating directions of magnetization then, according to G-K model the dependence $I_c(H)$ should exhibit rapid oscillations. Recently experimental observations of such oscillations and the critical current dependence on M-interlayer thickness have been shown [10- 12].

In order to obtain large proximity effect in superconducting structure with M-interlayer a relatively transparent S/AF interface is needed. That's why in-depth nvestigations of the interfaces composed of cuprate superconductors and antiferromagnetic cuprate are relevant [13, 14]. However, in spite of promising progress in fabrication of heterostructures with magnetic interlayer [10-16], there is still a lack of experimental results on Josephson junctions with antiferromagnetic interlayer, in particular with cuprate material. At the same time mutual influence of antiferromagnetism and the cuprate $d$-wave superconductivity at S/M interfaces in Josephson junctions is necessary to uncover.

In this paper we report on the experimental studies of the dc and rf current transport through superconductor/magnetic interfaces realized in hybrid Nb/Au/M/YBa$_2$Cu$_3$O$_{7-\delta}$ Mesa HeteroStructures (MHS) with the size in plane from 10×10 up to 50×50 μm$^2$. Here Nb is a conventional $s$-wave superconductor (S′), YBa$_2$Cu$_3$O$_{7-\delta}$ (YBCO) is cuprate superconductor with dominating $d$-wave order parameter (S$_d$), and Au is the normal metal. The M - interlayer is either the Ca$_{1-x}$Sr$_x$CuO$_2$ (CSCO) with x=0.15 or 0.5 which is in bulk a quasi-two dimensional Heisenberg antiferromagnetic cuprate [17, 18], or mixed-valence manganite La$_{1-y}$Ca$_y$MnO$_3$ exhibiting both the antiferromagnetism if y=0 and the ferromagnetism if y=0.3 [19]. We summarize recently obtained results on proximity effect and electron transport in hybrid heterostructure with M/S$_d$ interface taking into account the results published elsewhere [10, 12, 20].

Structure of experimental samples, the fabrication details and the measurements techniques are presented in Chapter 2. The X-ray diffraction data of the reference M-film and heterostructures, electrical and magnetic properties of M-interlayer films are discussed in Chapter 3. The superconducting transport in MHS is discussed in the Chapter 4 where dynamic behavior at microwave frequencies and the measurements of superconducting current-phase relation (CPR) are presented. The Chapter 5 describes the theoretical model based on multilayer magnetic structure consideration. It contains also comparison of the calculation results with the experiment. In particular, the magnetic field dependences of the superconducting current are compared with the Gorkov-Kresin model. Chapter 6 gives results for quasiparticle transport in MHS with different M/S$_d$ interfaces, and in Conclusions we summarize the obtained results.

2. **Experimental technique**

The double-layer epitaxial thin film structures M/YBCO were grown *in-situ* by pulsed laser ablation on (110) and (320)NdGaO$_3$ (NGO) substrates. Thus, the c-axis of the M/YBCO heterostructures is perpendicular to the substrate surface in the case of (110)NGO substrate or tilted from the substrate normal by the angle γ=11° for (320)NGO substrate. Typically, the d$_M$=10÷100 nm thick M- films were deposited on the top of 150 nm thick YBCO film. Doping of the CSCO films with Sr, and Ca or Sr of LaMnO$_3$ film was realized by varying the target composition. The M/YBCO heterostructures were covered *in-situ* by 10 nm thick Au film and later *ex-situ* 200 nm thick Nb film by DC-magnetron sputtering in Ar atmosphere. In order to fabricate Nb/Au/M/YBCO MHS we utilize optical photolithography, reactive plasma etching and the Ar ion-milling techniques. The SiO$_2$ protective layer was deposited by RF-magnetron sputtering and patterned afterwards in order to form the area of the MHS. An additional 200 nm thick Nb film was deposited on the top of the mesa area and patterned in order to form the superconducting wiring. Thus, the square S/N/M/S$_d$ MHSs having areas from 10×10 μm$^2$ and up to 50×50 μm$^2$ were fabricated (see Fig.1). For comparison a similar fabrication steps were used for structuring of the MHS without M-interlayer [21].

In order to avoid micro-shorts in the MHS several precautions have been made: (i) the deposited M films were thicker than the surface roughness of the YBCO layer[1]; (ii) Nb/Au bilayer is used as superconducting counter-electrode. Direct Nb deposition on top of the YBCO film results in formation of Nb/YBCO interface with very high resistance (~1 Ω×cm$^2$) due to Nb film oxidation. Thus, if the Au layer is locally damaged because of the finite surface roughness of the M/S$_d$ interface then niobium oxide is directly formed there, preventing micro-shorting of MHS. The 4-point measurement technique was used for electrical characterization of MHS: two contacts to the YBCO electrode and the two contacts to the Nb counter electrode (Fig. 1). The difference in electronic parameters of Au and M-layer determine the properties of potential barrier I$_b$ at the interface and it has the decisive contribution into the MHS resistance. Strong proximity effect in Nb/Au bilayer was confirmed for the MHS without M interlayer [21].

3. **Crystal structure and resistivity of the interlayer**
   **3.1. X-Ray diffraction analysis**

The XRD scans for CSCO (x=0.15) epitaxial films deposited on the (110)NGO substrate and on the YBCO/NGO heterostructure are presented in Fig. 2. The rocking curve measurements of the Full Width at Half Maximum (FWHM) of (002) peak of the autonomous CSCO film deposited directly onto NGO substrate revealed FWHM=0.07°. That value is smaller than FWHM=0.2° of (007) peak

---

[1] For M-interlayer with thickness 2-10 nm the NbBa$_2$Cu$_3$O$_x$ (NBCO) films were used instead of YBCO since it has smoother surface. Peak-to-valley surface roughness of single-layer (001)YBCO and M films measured by atomic force microscope were 2-5 nm, and 1-2 nm for the heterostructures based on NBCO.

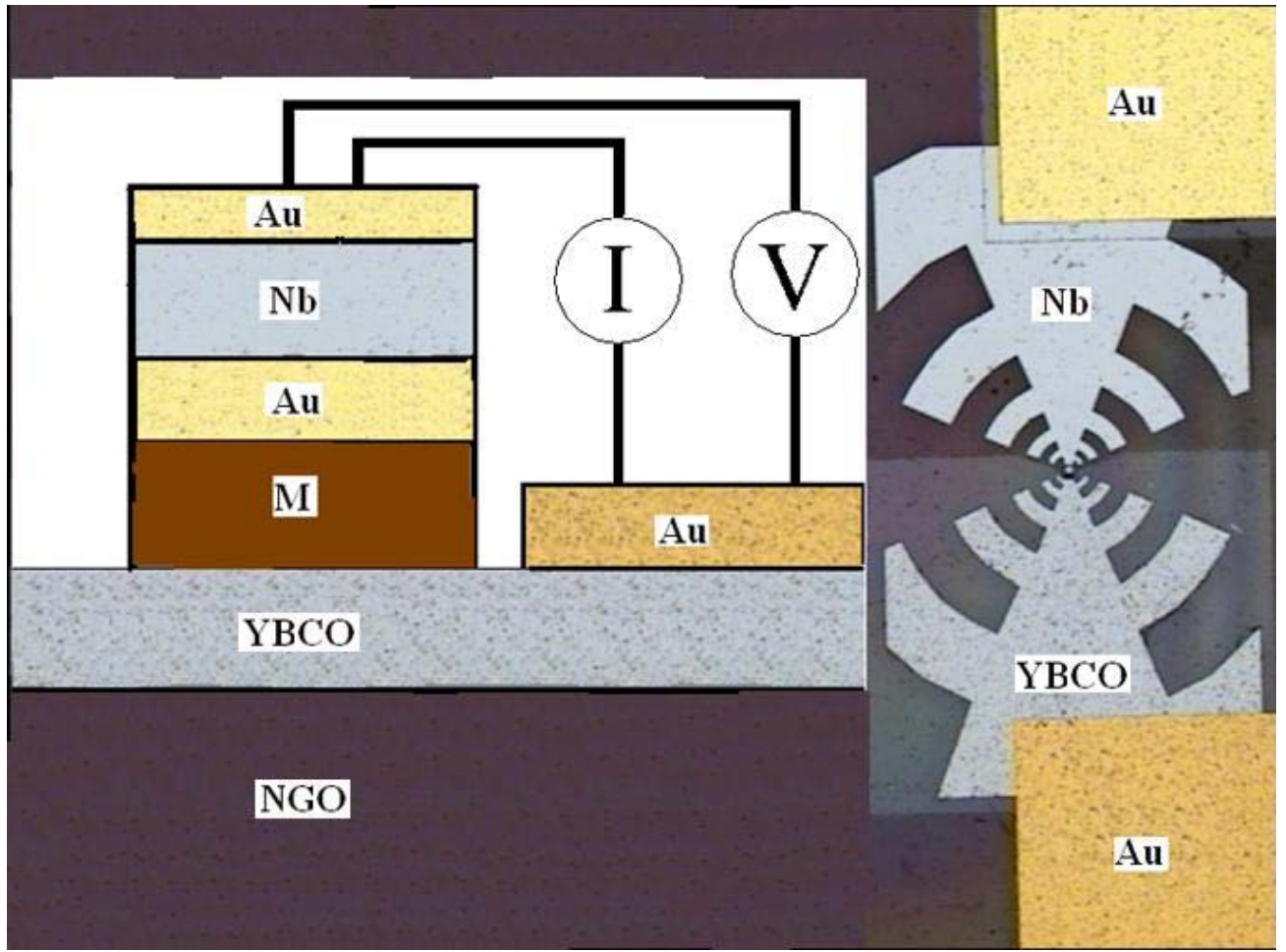

Fig.1. Crossection of the MHS with magnetic interlayer and circuit for current biasing is presented on left side. The layers thickness are as following YBCO - 200 nm, M-interlayer 5÷ 100 nm, Au- 10 ÷ 20 nm, Nb – 200 nm. Photo of MHS incorporated in log periodic antenna is presented on the right side

Table 1

| Structure | CSCO x=0.15 | CSCO/YBCO x=0.15 | | CSCO x=0.5 | CSCO/YBCO x=0.5 | | LMO | LMO/YBCO | |
|---|---|---|---|---|---|---|---|---|---|
| Reflection peak | (002) CSCO | (002) CSCO | (007) YBCO | (002) CSCO | (002) CSCO | (007) YBCO | (002) LMO | (002) LMO | (007) YBCO |
| $a_\perp$, nm | 0.321 | 0.322 | 1.169 | 0.333 | 0.336 | 1.177 | 0.398 | 0.397** | 1.171 |
| $\Delta\omega$, degree | 0.07 | 0.2* | 0.2* | 0.4 | 0.5* | 0.5* | 0.06 | 0.4 | 0.4 |

The crystal parameters for heterostructures and M-films of interlayer.*The estimation of Δω were made from 2θ/ω scan measured for the best YBa$_2$Cu$_3$O$_x$ film. The rocking curve measurement of the single crystal substrate (110)NGO showed FWHM=0.006° determined by the resolution of X-ray diffractometer [22]. The FWHM values of the rocking curve of the CSCO film deposited on YBCO films are increased by several times. Similar behavior is observed for CSCO (x=0.5) and for LMO films deposited on YBCO/NGO heterostructure (see table 1). All M-films deposited over the YBCO/NGO demonstrate a broadening of the rocking curve, manifested in a reduction of crystallographic quality and in a minor change of the lattice constants. According to [23-24] the interfaces in cuprate/manganite might be coherent, free of defects, exhibiting roughness less than 1 nm and there is no major chemical ion intermixing. However X-ray absorption spectroscopy analysis of the interface show some cation displacement [25].

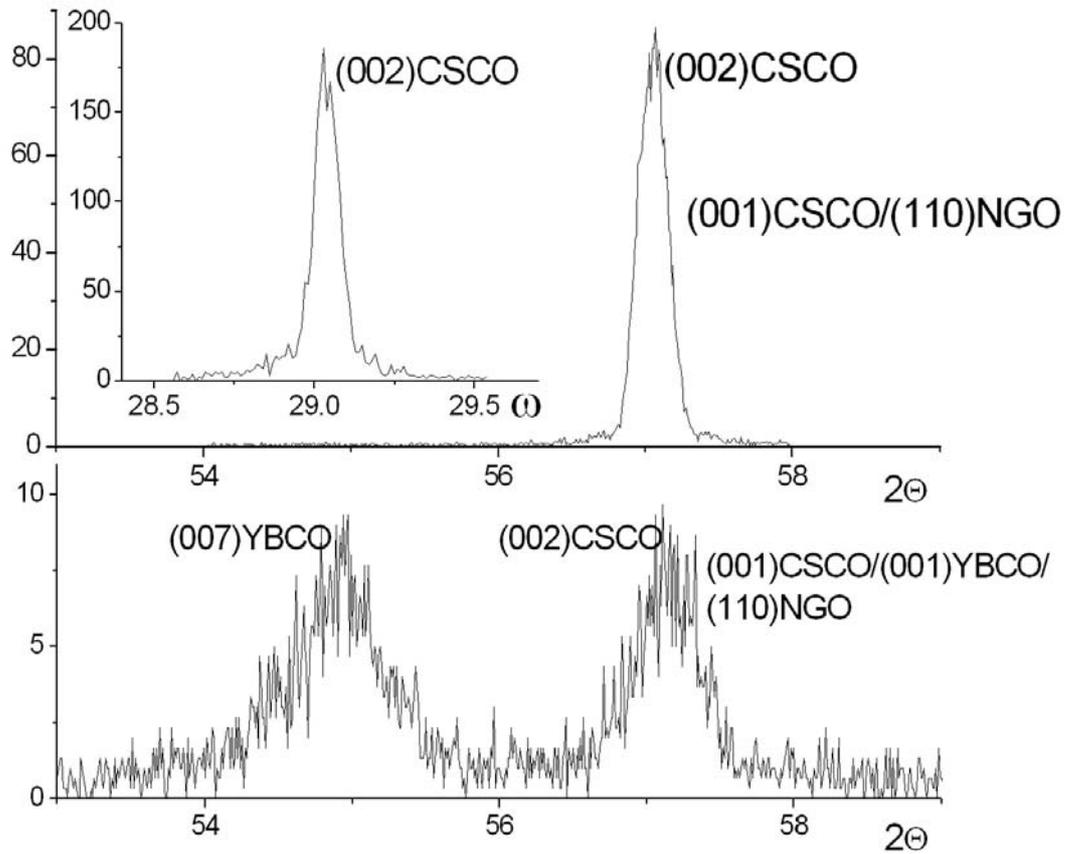

Fig.2. X-ray θ-2θ scans for an epitaxial (001)CSCO film (d=50 nm) on the (110)NGO substrate (the upper graph) and (001)CSCO/(001)YBCO/(110)NGO thin film multilayer structure (d=100 nm) (the lower graph). The rocking curve of (001)CSCO/(110)NGO bilayer is shown in the inset

**3.2. M-films resisivity**

Temperature dependences of resistivity of the autonomous M-films are presented in Fig. 3. Increase of the resistivity is observed with decrease of temperature for all M-films excluding the LCMO (y = 0.3) film which show metal-insulator transition at T = $T_{cu}$ = 250 K. Note, that LCMO films have lower resistance than other M-films at room temperature.

The temperature dependences of conductivity of CSCO, LMO and LCMO (for T>$T_{cu}$) films well enough correspond to the three-dimensional hopping model (Fig.4):

$$\rho = \rho_0 \exp((T_0/T)^{1/4}) \quad , \quad (1)$$

where $\rho_0$ and $T_0 = c\,e\,/\,(k\,N(E_f)\,\lambda^3)$ are characteristics of hopping conductivity (c - constant, k - Boltzmann constant, N($E_f$) - the density of states at the Fermi level, $\lambda$ - length of localization). The degree of exponential temperature dependence is determined by the dimension of the system and in our case corresponds to the three-dimensional variable range hopping conductivity. We have estimated the length $\lambda$ and $T_0$ from experimental data assuming $N(E_f) = 10^{20} cm^{-3} eV^{-1}$ and $c \approx 18$: For CSCO[2] $\lambda = c\, (e/k_B\, N(E_f)T_0)^{1/3} \approx 1$ nm at $T_0 = 3\,10^6\,°K$, and at $T = 300$ K the activation energy $E_{hop} = k_B T (T_0/T)^{1/4} = 0.26$ eV is somewhat smaller than the gap, defined from optical measurements [26]. Unlike the conventional insulators with activation type of conductivity where dielectric gap is caused by crystal field, the band gap in Mott insulators is formed due to strong electronic correlations. Mott insulators demonstrate rise of conductivity with decrease of temperature even in the case of half-filled conductance band. No activation type dependence was observed for M-film resistance even at $\rho > 10^3\,\Omega$ cm.

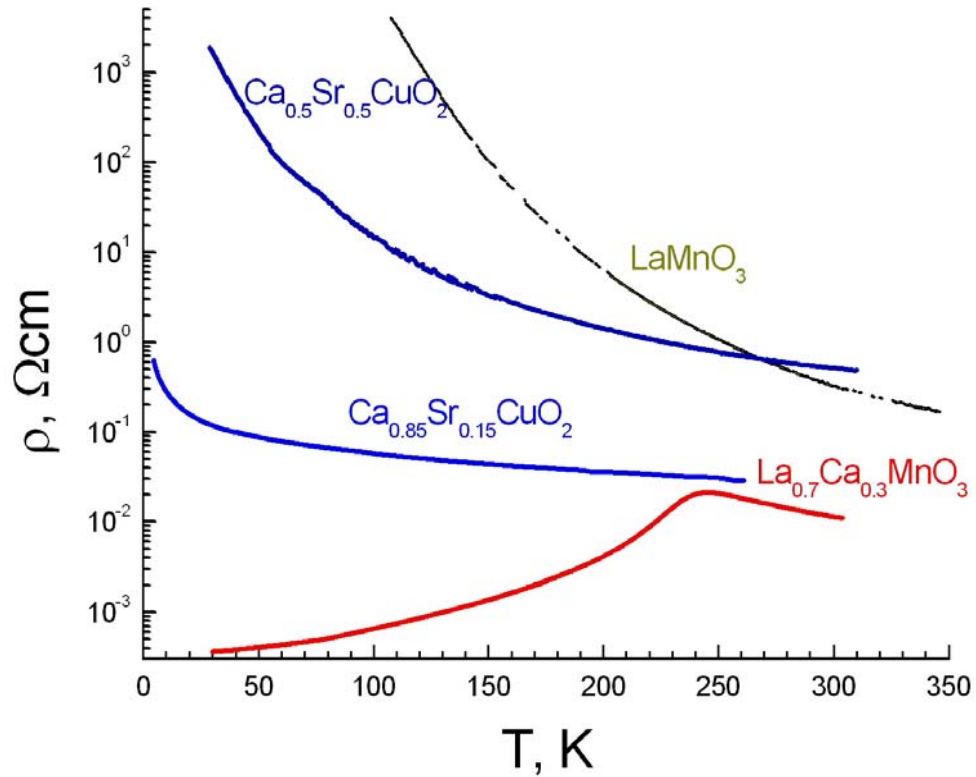

Fig.3. Temperature dependence of autonomous (deposited directly on the substrate) M-films. The thickness of M-film are order of 80 nm

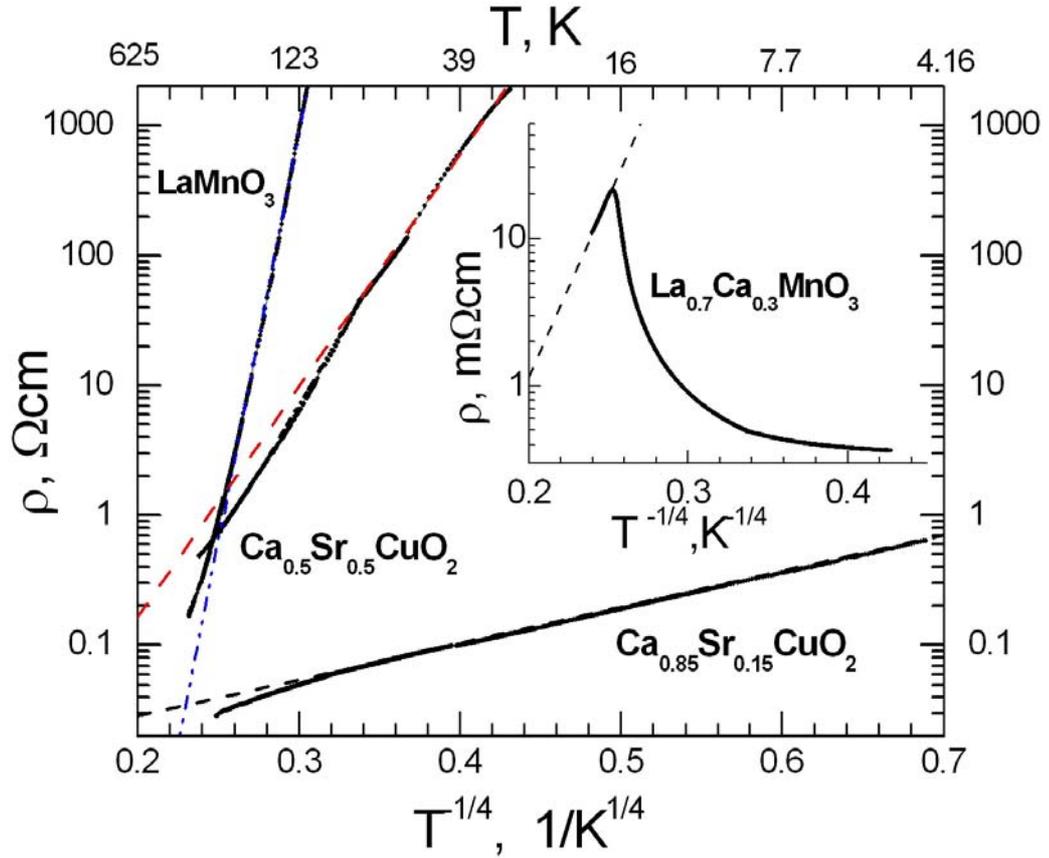

Fig. 4. Temperature dependence of autonomous M-films in semilog scale. 3D hoping conductivity shown by dash line is clearly seen as in (1). $\rho(T)$ for LCMO film is shown on inset.

---

[2] Here and further the CSCO with x=0.5 will be discussed.

### 3.3. Magnetic properties of M-films

Magnetic sublattice of M-films is defined by crystal structure. It is known that magnetic sublattice of CSCO and LMO corresponds to the G- and the A-type AF ordering, respectively (Fig. 5) [17-19]. In order to determine the ferromagnetic $T_{cu}$ or antiferromagnitic transition temperature ($T_N$) the magnetic resonance spectra were studied in the temperature range T= 80 – 300 K using Electron Paramagnetic Resonance (EPR) spectrometer. Concentration of the paramagnetic centers was determined from comparison of the EPR spectrum line with the $Mn^{2+}$ of the standard MgO: placing the Mn sample in the same cryostat. However, we were not able to resolve the paramagnetic resonance line of CSCO films with EPR technique approach. Moreover, usage of much sensitive SQUID meter also didn't help to determine the Neel temperature ($T_N$) because of too high resonant moments of the Nd ions in NGO substrate used for the deposition of CSCO films. The results of neutron study [17, 18] obtained on polycrystalline samples gave $T_N = 450$ K. Therefore,

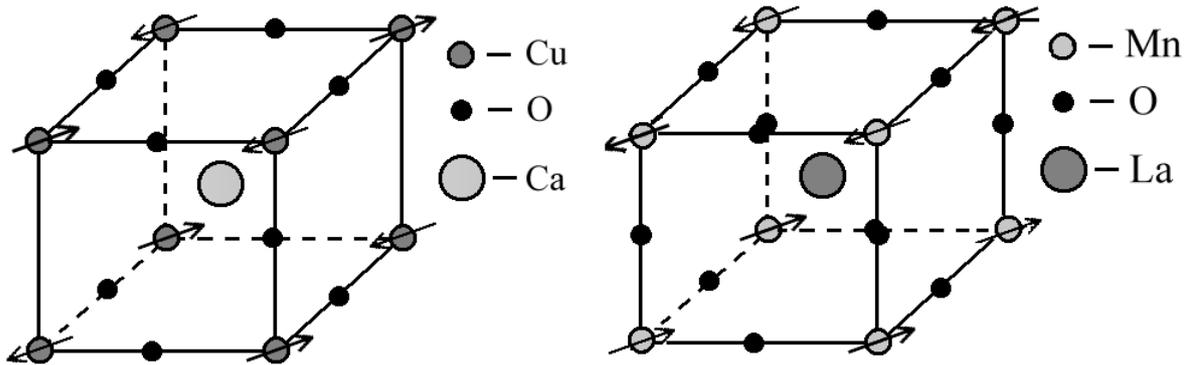

Fig.5. Magnetic sublatice structure for CSCO(left side) and LMO(right side) interlayer. Internal layer magnetization for G-type and A-type AF-ordering are shown schematically

we assumed that the CSCO films are in G-type antiferromagnetic Mott insulator state at temperatures 4.2 – 40 K where the electrical measurements have been carried out.

A ferromagnetic resonance (FMR) for LCMO films (down to thickness of 10 nm) was observed with temperature decrease. Change in the slope of FMR magnetic field $H_0$ with T defines the Curie temperature of ferromagnetic transition $T_{cu}=180$ K (see Fig.6). That is close to the temperature of metal-insulator transition determined from the temperature dependence of the LCMO film resistance. Effect of metal-insulator transition in the LCMO film is manifested in the temperature dependence of the MHS (see Chapter 6).

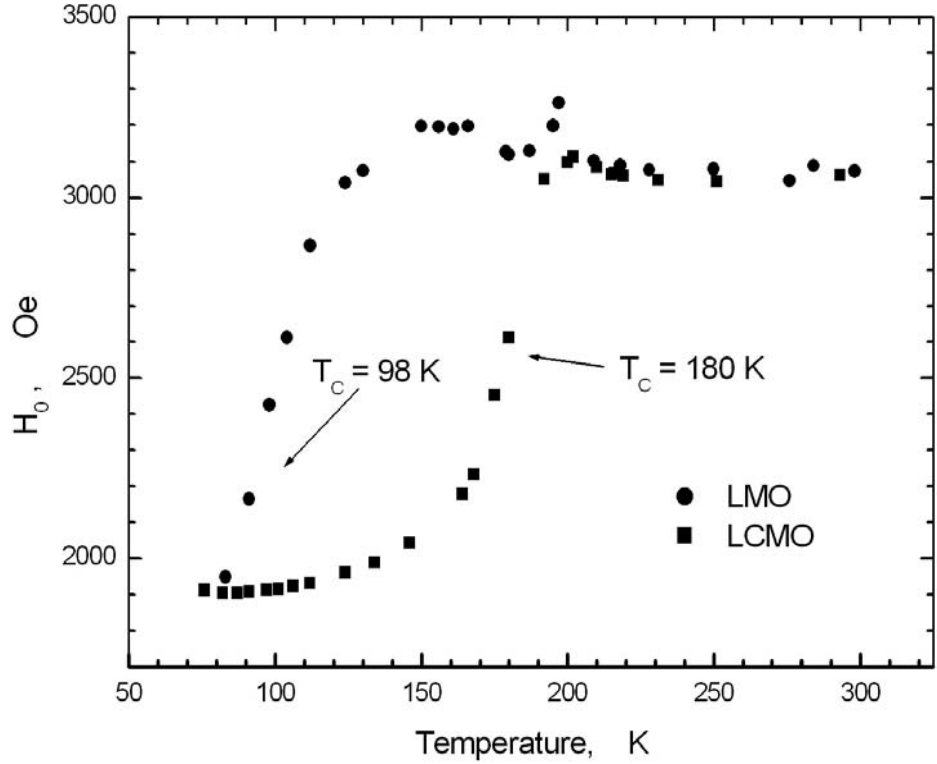

Fig.6. The temperature dependence of FMR magnetic filed $H_0$ for autonomous LMO (filled circles) and LCMO films (filled squires). The Curie temperature ($T_{cu}$) is defined by the change in the slope of the ferromagnetic resonance field. The thickness of the film is smaller then for data presented on Fig.3 results in reduction of $T_{cu}$.

The temperature dependencies of resistivity of LMO films do not demonstrate resistive metal-insulator transition (see Fig.3). However the FMR spectra shows ferromagnetic resonance with the Curie temperature $T_{cu}$ =98 K for the both kind of films deposited on the pure substrates and on the top of the YBCO/NGO heterostructures. It is commonly accepted that the double –exchange interaction between $Mn^{3+}$ and $Mn^{4+}$ ions are responsible for ferromagnetism in highly doped manganites similar to the investigated LCMO [19]. For low-doped compositions the superexchange interaction between $Mn^{3+}$ ions is responsible for appearance of FM and AFM phases. Jahn-Teller distortion plays important role in ordering of $Mn^{3+}$ ions [27, 28]. Ferromagnetism is observed both in the low doped $La_{1-x}Mn_{1-x}O_3$ compounds and the $LaMnO_{3+\delta}$ with an oxygen nonstoichiometry. The strain of the manganite films due to the influence of the substrate [29] can enhance the ferromagnetism in LMO similar to ferromagnetism induced by external pressure [27].

4. **Superconducting transport in heterostructure with M/S interface**

### 4.1. Temperature dependence of hybrid heterostructure resistance

The resistance of MHS is the sum of resistances

$$R = R_{YBCO} + R_M' + R_{M/Y} + R_b + R_{Nb} \qquad (2),$$

of YBCO electrode ($R_{YBCO}$), resistance of M-interlayer ($R_M'$), M/YBCO interface resistance ($R_{M/Y}$), Au/M interface resistance ($R_b$), and resistance of the Nb electrode ($R_{Nb}$). The contribution of the thin Au film can be neglected [21]. At temperatures T higher than superconducting critical temperature of the YBCO film ($T_c$) the temperature dependence of the MHS $R(T)$ is similar to that of the autonomous YBCO film (linear decrease with T lowering) as seen in Fig.7. For samples with thick CSCO interlayer ($d_M$>100 nm) or LMO(LCMO) (even for 10 nm thick) a large deviation from linear decrease R(T) was observed[3]. At $T_c' < T < T_c$ (where $T_c'$ is the critical temperature of the Nb/Au bilayer) the MHS resistance is determined by the interfaces M/YBCO, Au/M, Nb/Au, Nb wiring, and the resistance of the M-interlayer $R_{M'}$. Independently measured characteristic resistance of the Nb/Au interface (~$10^{-12}$ Ω·cm$^2$) results in ~1μΩ - a negligibly small contribution to resistance of MHS [21].

Taking into account the epitaxial growth of CSCO / YBCO structure, and the similar values of the Fermi velocities in contacting materials, one can assume that resistance of the boundary $R_{M/Y}$ is small compared to the resistance of the Au/CSCO interface, for which the difference between Fermi velocities of Au and the CSCO is significant [21]. Resistance measurements of a autonomous CSCO films (x = 0.5) deposited on NGO substrate show resistivity $\rho$=10$^3$-10$^4$ Ω·cm at low temperatures (see Fig.3) resulting in the resistance of the MHS more than 1 kΩ.

However, for MHS with thin CSCO interlayer $d_M$ <50 nm no increase of resistance was observed. Compare with autonomous CSCO films (curve 3 Fig.7) the resistance of MHS (curves 1 and 2 Fig.3) weakly dependents on temperature and is significantly smaller than the resistance of the CSCO interlayer $R_M'$ obtained from measurements of the autonomous CSCO film (Fig.3).

Characteristic resistance $R_NA$ of the MHS (A = L$^2$ - MHS area) measured at T = 4.2 K exponentially increases with d$_M$ (see Fig.8). Note, if the main contribution to $R_NA$ comes from the interlayer, then a linear increase of $R_NA$ with d$_M$ should be observed.

---

[3]The detail analysis of $R(T)$ behavior for LCMO (LMO) interlayer will be given in Chapter 6.

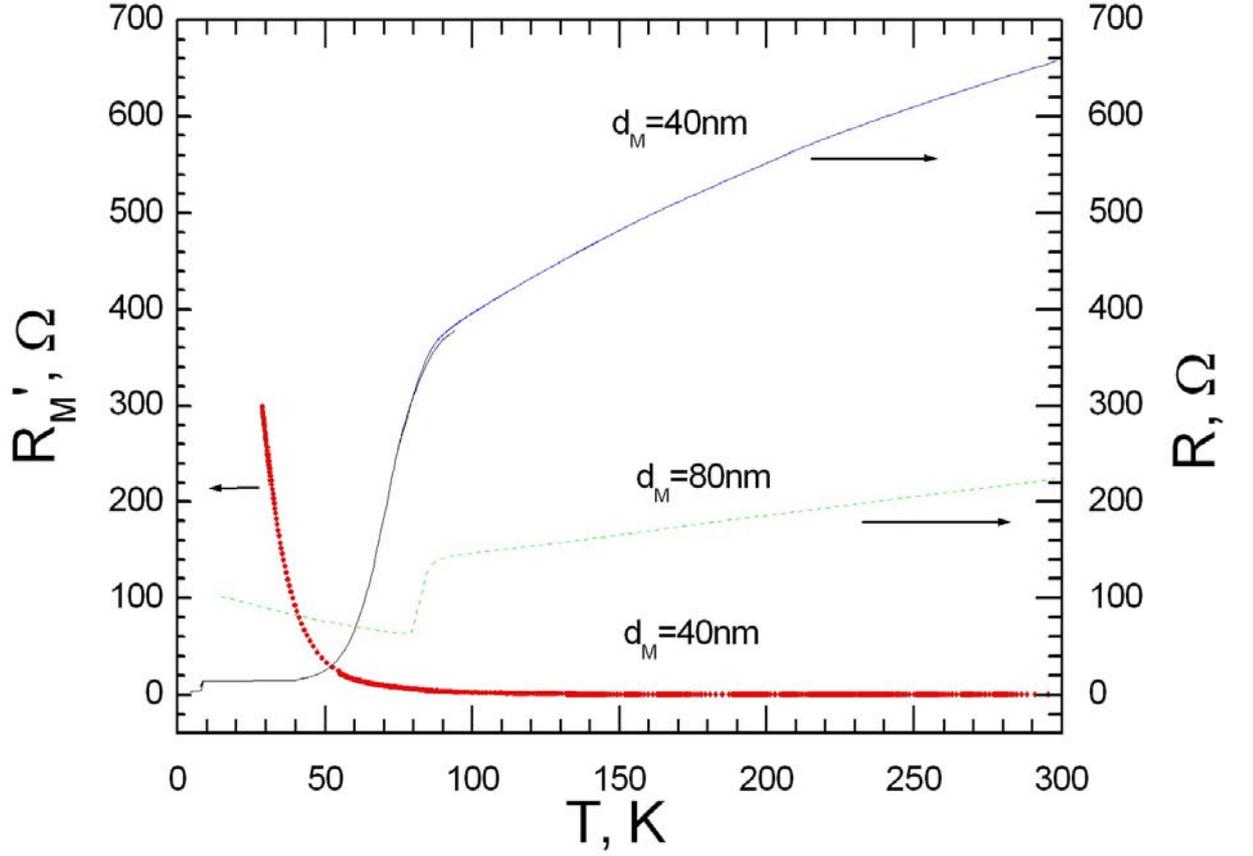

Fig.7. Temperature dependence of resistance for MHS with two thickness of CSCO-interlayer: $d_M$=40 and 80 nm. The dependence of interlayer $R_M' = \rho_M/d_M$ for $d_M$=40 nm where $\rho_M$ is calculated from the resistance for autonomous film (Fig.3) is shown by dashed line.

The exponential dependence of the $R_N A(d_M)$ can be explained by a changes of conductivity of the CSCO interlayer due to oxygen nonstoichiometry in thin $d_M$ <80 nm interlayer film caused by a rearrangement of the electronic subsystem [30-32]. Namely, the non-uniform electron doping across the interlayer thickness can explain the exponential dependence of the $R_N A (d_M)$ because of large Au/CSCO interface contribution to the resistance of the boundary determined by the ratio of Fermi velocities of the contacting materials [21]. Experimental $R_N A(d_M)$ data show typical dependence for diffusion processes when doping level in CSCO toward to the Au/CSCO interface depends exponentially on the distance from the doping source. Electrical conductivity of MHS at high voltages (up to 100 mV) differs from conductivity known for ballistic superconducting contacts or superconductor-normal metal-superconductor junctions, but looks like tunneling type. This feature presumably is due to the presence of low transparent Au/CSCO interface. As it was shown in [13], despite of a weak cation diffusion at the interfaces of the cuprates (about 1-2 atomic units), the occurrence of superconductivity in the metal-insulator transition in cuprates may be caused by electronic rearrangement or oxygen nonstoichiometry as it may happen at the interface of strongly correlated Mott and band insulator [30]. Such charge transfer processes may lead to a significant alteration of the electronic subsystem. At the same time, we do not rule out the oxygen nonstoichiometry at the interface similarly to the case discussed in [14]. It may cause the transition of thin ($d_M$ <50 nm) CSCO layers into the metallic state, as was observed in [31] with decreasing the oxygen content in the CSCO films during their growth. A significant change in the electronic conductivity can occur due to interaction between the oxygen atoms in CSCO/YBCO interface leading to the observed dependence of $R(T)$ in the temperature range $T < T_c$. Thus at $d_M$ <70 nm the resistances of our MHS are determined mainly by the Au/CSCO interface and the contribution of the CSCO layer resistance is important only at $d_M>$ 70 nm when critical current is strongly suppressed [4].

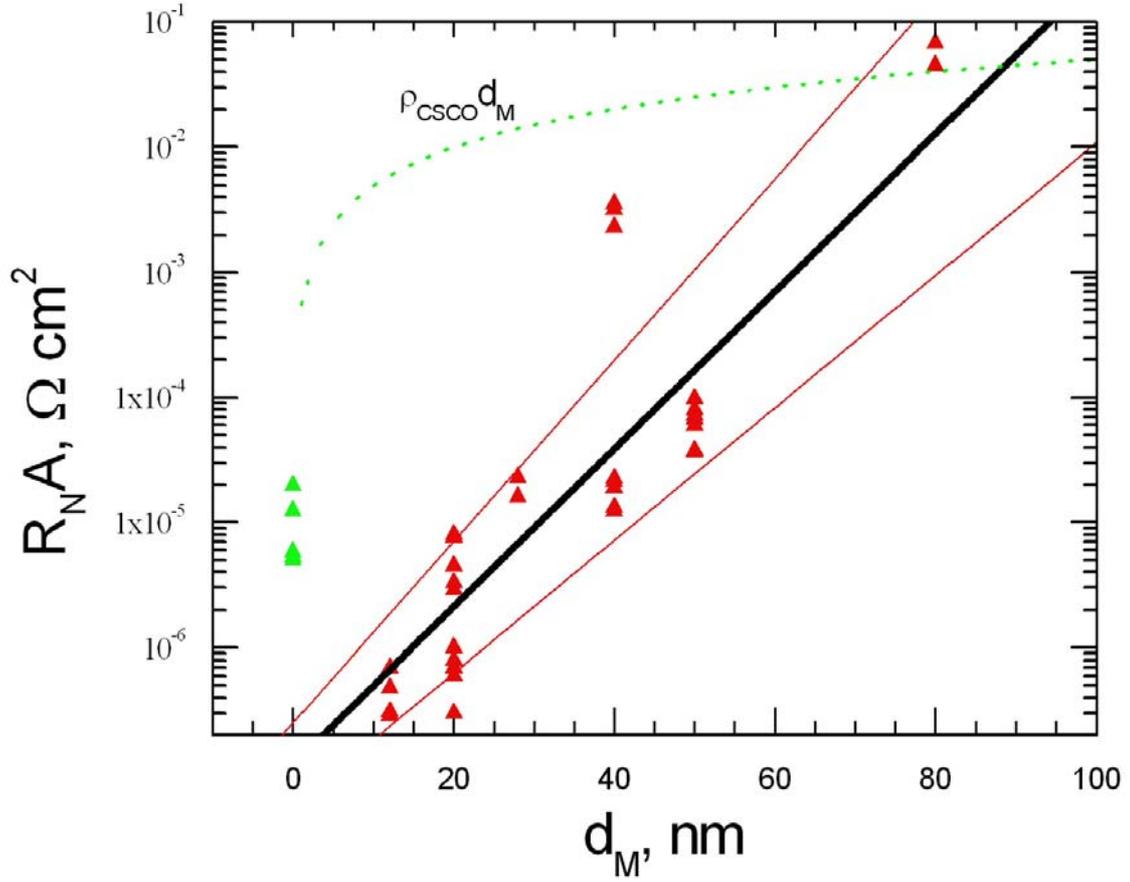

Fig.8. Interlayer thickness ($d_M$) dependence of characteristic resistance of the MHS ($R_N A$) with CSCO interlayer. The exponential increase of the resistance with decay length 6.3 ±0.8 nm is shown by solid line. Thin lines determine the deviation interval for experimental points. The dashed line is $\rho_{CSCO} d_M$ for $\rho_{CSCO}$ which is calculated from Fig.3

### 4.2 Superconducting current

The superconducting current ($I_C$) is clearly observed at $T$ = 4.2 K for MHS with CSCO interlayer with thickness $d_M$ =10-50 nm (see table 2). Data in table 2 show that the $I_C R_N$ products are several times higher than $I_C R_N$ of MHS without M-interlayer. The characteristic resistance $R_N A$ slightly varies with thickness $d_M$ and Sr doping level x.

**Table 2**

| N | x | $d_M$, nm | $L$, μm | $I_C$, μA | $R_N$, Ω | $V_C$, μV | q |
|---|------|-----------|---------|-----------|----------|-----------|------|
| 1 | 0.15 | 50 | 10 | 44 | 3.0 | 132 | 0.2 |
| 2 | 0.15 | 20 | 10 | 49 | 1.9 | 93 | 0.08 |
| 3 | 0.5 | 20 | 10 | 334 | 0.71 | 237 | 0.4 |
| 4 | 0.5 | 50 | 10 | 2.5 | 60 | 150 | 0.13 |
| 5 | - | 0 | 40 | 160 | 0.36 | 58 | 0.5 |
| 6 | - | 0 | 20 | 18 | 3.6 | 65 | 0.4 |

DC parameters of MHS with CSCO interlayer measured at T=4.2K. DC parameter for MHS with LMO and LCMO interlayer and also without M-interlayer are presented for comparison. x is doping level of CSCO, $d_M$ is thickness of M-interlayer, L is linear size of MHS, $I_c$ is critical current, $R_N$, is normal resistance, $V_c=I_c R_N$, q is the ration of second and first harmonics of superconducting current

The MHS can be considered as S'/$I_b$/M/$S_d$ structures where $S_d$ is YBCO electrode with dominant d-wave order parameter and a small admixture of s-wave component $\Delta(\theta)=\Delta_d\cos2\theta+\Delta_S$, $\Delta_d$ and $\Delta_s$ are amplitudes of d-wave and s-wave superconducting pair potential, S' is superconducting Nb/Au billayer (due to strong proximity effect in thin Au), $I_b$ is the barrier formed due to defects, nonstichiometry at the Au/CSCO interface and the difference of Fermi velocities for contacting materials [21]. . The variation of superconducting pair potentials $\Delta_d$ and $\Delta_S$, in the vicinity of the barrier ($I_b$) due to proximity effect at M/$S_d$ interface should be taken into account. As follows from [35] $\Delta_d$ could rapidly decrease and $\Delta_S$ increase at the M/$S_d$ interface.

The $I_c(T)$ curves follow the temperature dependence of the Nb superconducting gap similar to MHS without M-interlayer [10, 21]. Note, such behavior also follows from calculations [8] of long-range proximity effect in M-interlayer. However, we did not observe neither any noticeable change of $I_cR_N$-product from the thickness of the CSCO layer, nor a square-law increase of critical current with decreasing temperature at $T$ near $T_c'$. Note, all of investigated MHS with thickness $d_M$ of tens nanometers had critical current $I_C$>1 µA, thus the penetration depth of superconducting correlations in CSCO significantly exceeds the coherence length of a polycrystalline AF layers of FeMn [9].

---
[4] Temperature dependence of MHS with manganite interlayer will be discussed in 6.1.

### 4.3. Current phase relation of superconducting current

The dominant d-wave symmetry of the superconducting order parameter in $S_d$ electrode may result in non-sinusoidal CPR for MHS with c-axis orientedYBCO, which contains first ($I_{c1}$), second ($I_{c2}$), etc. harmonics, i.e. $I_s(\varphi)=I_{c1}sin\varphi+I_{c2}sin2\varphi+\ldots$. Note, the d-wave component $\Delta_d cos2\theta$ may give a second harmonic in CPR ($I_{c2}R_N$) proportional to the next order of the $I_b$ transparency in the structure [33, 34, 36-38].

The values of the second harmonics of the CPR were defined from measurements of Shapiro steps, which arise on the I-V curves irradiated by microwaves at experimental frequencies 36-120 GHz. All MHS under applied monochromatic mm-wave radiation had very well oscillating with power Shapiro steps on I-V curves (inset to Fig.9). Oscillations of the Shapiro steps amplitudes *vs.* applied microwave power (Fig.9) confirm the Josephson effect origin of the superconducting current. Less than 20% difference has been observed between the critical frequency $f_c=2eV_c/h$=71 GHz calculated from $V_c=I_cR_N$=147µV (static estimation of $f_c$) and the $f_c$=56 GHz determined from the maximum value of the first Shapiro step using the Resistive Shunted Junction (RSJ) model approach (dynamic $f_c$) [39]. The correspondence between these two values clearly indicates the absence of micro-shorts [16]. The deviation from the RSJ model becomes smaller if we take into account a presence of the second harmonic component in CPR which is manifested by fractional Shapiro steps on the I-V curve (inset Fig.9) observed at all experimental frequencies up to 120 GHz. It is known, that fractional Shapiro steps may originate also from the finite capacitance of the Josephson junctions (McCumber parameter $\beta_c=4\pi eI_cR_N^2C/h$>1) [21, 39]. We calculated values of $\beta_c$=2-6 from the hysteretic I-V curves of MHS. In order to investigate the influence of the second harmonic in CPR and the capacitance $C$ on dynamic properties we have studied dependences of the critical current $I_C(a)$ and the first Shapiro step $I_1(a)$ on the normalized amplitude of electromagnetic radiation (Fig.9). The performed calculations of the amplitudes of the Shapiro steps based on the modified RSJ model show that at $hf_e>2eI_cR_N$ the contribution of the MHS capacitance is relatively small and the $I_C(a)$ and $I_1(a)$ dependences are determined mainly by the second harmonic of the CPR. The experimental data presented in Fig. 9 are well fitted to the theoretical dependencies [21] calculated taking into account the amplitude $I_{c2}$ of the second harmonic in the CPR $q=I_{c2}/I_c$=0.2. Note, the sign of $q$ can be determined by fitting the dependence of half integer Shapiro steps $I_{1/2}(a)$ to the corresponding theoretical dependence [21]. This procedure gives us negative $q$ <0. Taking $q$>0 the calculated maximum of $I_{1/2}(a)$ becomes much higher than the measured one. Negative sign of second harmonic in CPR is native for superconducting junctions with d-wave order parameter [34, 36-38]. The presence of second harmonic in CPR of MHS indicates an existence of small, but nonzero d-wave component of pair potential on S'/M interface.

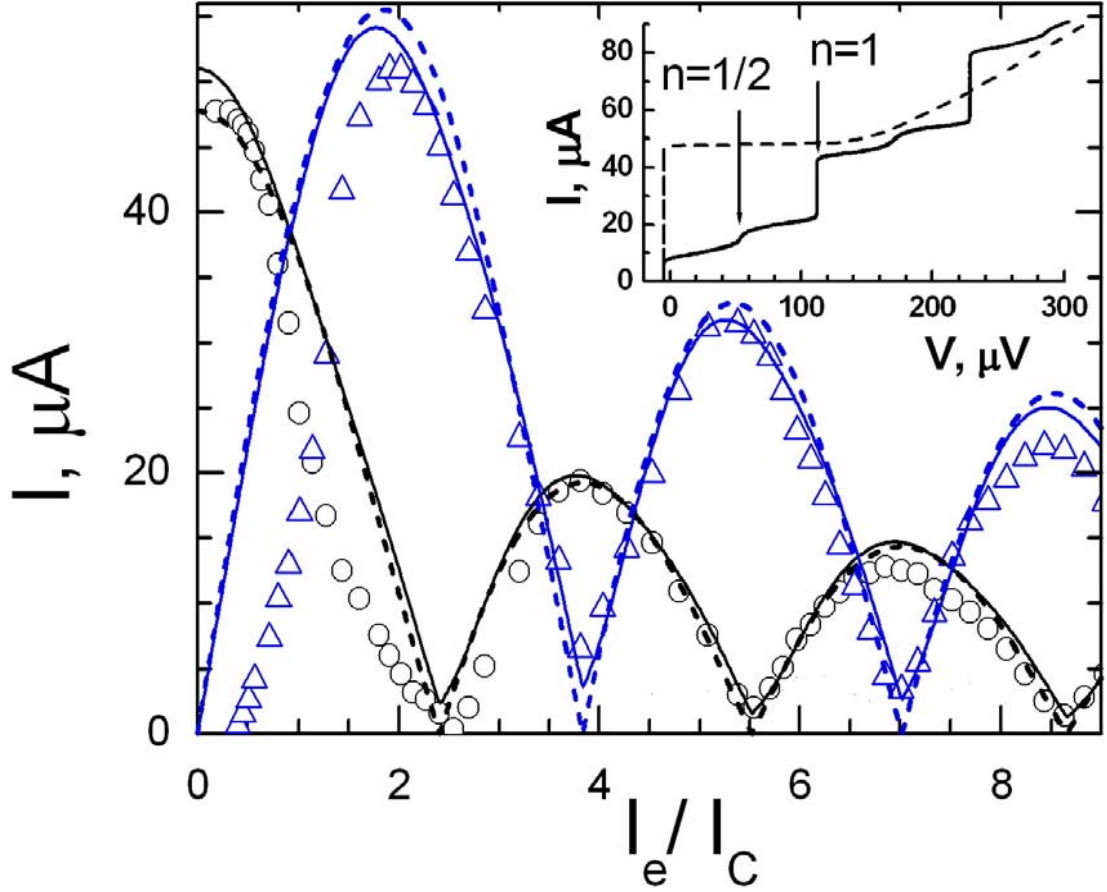

Fig. 9. The experimental critical $I_c(I_e)$ (circles) and first Shapiro step $I_1(I_e)$ (triangles) dependences on external microwave current $I_e$ for Nb/Au/CSCO/YBCO MHS for microwave frequency $f_e$=56 GHz and $T$=4.2 K. The solid and dashed lines correspond to the $I_c(I_e)$ and $I_1(I_e)$ theoretical curves numerically calculated from the modified RSJ model taking into account the second harmonic in the superconducting current-phase relation $q=I_{c2}/I_c$=0.2 and $q$=0 correspondingly[21]. Autonomous (solid line) and I-V curves of the MHS under influence of the external microwave current (dash line) are shown in inset. Positions of one integer $V_1=nhf_e/2e$ ($n$=1) and half-integer n=1/2 Shapiro steps are denoted.

## 5. Proximity effect on AF/S interface.

### 5.1 The model of multilayered magnetic structure.

In order to describe the experimental results we consider a model of S'/I$_b$/M/S structure, where I$_b$ is a barrier with the low transparency and M– interlayer located between two s-wave superconductors S' and S (inset to Fig.10). As a model of M-interlayer we consider the multilayer magnetic structure (MMS- model), i.e. it is assumed that M interlayer consists of N ferromagnetic layers each one with the thickness $d = d_M / N$ much larger than interatomic distance. The value of the layer exchange energy $J_{ex}$ is assumed to be small compared with the Fermi energy. Magnetizations of layers are assumed to be collinear and have orientation in plane of S/M interface. The considered model enables us to use approach based on quasiclassical Green's function equations (see, e.g., [1, 2][5])

We also assume that condensate Green's function in the M-interlayer is small that is realized for small transparency of M/S interface or for arbitrary transparency of the interface at the temperatures close to $T_c$. We assume that the transparency of the I$_b$ barrier at S'/M interface is smaller than for M/S interface and the influence of the superconductor S' on the condensate function in the layer M can be neglected. Let us analyze the dirty limit, where the mean free path $l$ is small compared with $d$ and $\tau J_{ex} \ll 1$, where $\tau$ is the impurity scattering time. In this case the isotropic part of the Green's function $\langle f_\sigma \rangle = s_\sigma \gg f_\sigma - s_\sigma$, ($s_\sigma \ll 1$) satisfies the wave type equation,

$$\partial_{xx}^2 s_\sigma - k_\sigma^2 s_\sigma = 0, \qquad (3)$$

where $k_\sigma = [2(\omega + i\sigma J_{ex}(x))/D]^{1/2}$, $\omega = \pi T(2m+1)$ is the Matsubara frequency (we assume that $\omega > 0$), $D = v_F l/3$ is the diffusion constant in the F-layers. Solutions for $s_\sigma$ at $0 < x < d_M = Nd$ in the case of antiferromagnetic ordering of the magnetization in the layers can be represented as follows:

$$s_\sigma(x) = \begin{cases} \cdots \\ A_n^\sigma \cosh k_\sigma(x - x_{n-1}) + B_n^\sigma \sinh k_\sigma(x - x_{n-1}), & x_{n-1} < x < x_n \\ A_{n+1}^\sigma \cosh k_{-\sigma}(x - x_n) + B_{n+1}^\sigma \sinh k_{-\sigma}(x - x_n), & x_n < x < x_{n+1} \\ \cdots \end{cases} \qquad (4)$$

where $x_n = nd$. Taking into account the continuity of the function $s_\sigma$ and its derivative at $x = x_n$, one can obtain the following recurrence relations for the coefficients

$$\begin{pmatrix} A_{n+1}^\sigma \\ B_{n+1}^\sigma \end{pmatrix} = \mathbf{M}_{\sigma_n} \begin{pmatrix} A_n^\sigma \\ B_n^\sigma \end{pmatrix} \qquad (5)$$

in which the matrix $\mathbf{M}_\sigma$ is determined by the expression

---

[5] In this section we assume the Boltzmann's and Planck's constants equal to 1

$$\mathbf{M}_\sigma = \begin{pmatrix} 1 & 0 \\ 0 & q_\sigma \end{pmatrix} \begin{pmatrix} \cosh \lambda_\sigma & \sinh \lambda_\sigma \\ \sinh \lambda_\sigma & \cosh \lambda_\sigma \end{pmatrix} \quad (6)$$

where $q_\sigma = [(\omega + i\sigma J_{ex})/(\omega - i\sigma J_{ex})]^{1/2}$, $\lambda_\sigma = k_\sigma d$, $\sigma_n = (-1)^{n+1}$. Taking into account the boundary condition at $x = 0$, $\partial_x s_\sigma(0) = 0$, we get $B_1^\sigma = 0$. Therefore, using the notation

$$\prod_{n=1}^{N-1} \mathbf{M}_{\sigma_n} = \|m_\sigma^{ij}\| \quad (7)$$

For boundary conditions $x = 0$, we obtain the relationship for the condensate function at different edges of the M-layer:

$$s_\sigma(0) = c_\sigma s_\sigma(d_M), \quad (8)$$

where

$$c_\sigma = \frac{1}{m_{\sigma N}^{11} \cosh \lambda_{\sigma N} + m_{\sigma N}^{21} \sinh \lambda_{\sigma N}}$$

$\sigma N = (-1)^{N+1}\sigma$. The condensate Green's function $c_\sigma$ describe the evolution of the condensate function in the M-interlayer for AF ordering of magnetizations in the F-layers. The analysis of this function shows that, provided the thickness of layers $d \ll \xi_{ex}$, $\xi_{ex} = (D/\pi J_{ex})^{1/2}$ for this ordering the condensate Green's function can penetrate into the M-layer over the depth of the order $\xi_N = (D/\pi T)^{1/2}$. In other words, under the condition $d \ll \xi_{ex}$, the "long-range proximity effect" (LRPE) may take place for AF ordering of the magnetization. Note that unlike the predicted in [2, 40] LRPE associated with the triplet component of the condensate Green's function appearing under the condition of spatially non-homogeneous non-collinear magnetization in ferromagnetic, considered here the LRPE is associated with the singlet component of the condensate Green's function. The origin of LPRE is related with the weakening of the pair-breaking effect due to exchange field effective averaging and with the additional scattering of Cooper pairs at the inter-layer interfaces. The LRPE and other features of proximity effect in structures containing multi-layer magnetic structures with AF ordering of the magnetization is realized also in clean limit when thickness $d_M$ exceeds the mean free path l. The exponential decay of the critical current is characterized by $\xi_N = hv_F/4\pi kT$ in that case [8, 41- 44].

### 5.2 Thickness dependence

Fig. 10 shows experimental and the calculated dependences of superconducting current density $j_c$ on normalized thickness $d_M$ of M-layer for different level of exchange field $J_{ex}/\pi kT$. Exponential decrease of $j_c$ with $d_M$ can be seen for both the experimentally measured points and the curved calculated within MMS-model. For chosen fitting parameters N=20 and $J_{ex}/\pi kT$ =2 the decay depth corresponds to $\xi_{AF}$ =7 ± 1 nm. Thus, in the CSCO interlayer $\xi_{AF}$ significantly exceeds the coherence length of the polycrystalline metallic AF interlayer FeMn [9]. Statistical analysis of experimental data obtained from the $j_c(d_M)$ dependences of the all of investigated samples also gives close values of $\xi_{AF}$. Theoretical dependences have been calculated for identical superconductors S and S', but qualitatively results remain unchanged for different superconductors. In experiment the superconductors were not identical, moreover in YBCO electrode the s-symmetry is not the main in the wave potential.

The theoretical curves demonstrate decrease of $j_c$ with the exchange energy $J_e$. A different behavior for $j_c$ is seen for the structures with the even and the odd number N of layers. Assuming that quasiparticle mean free path of $l$~10 nm in the CSCO film (approximately equal to the thickness of interlayer), we evaluate the value of the Fermi momentum $p_F=\hbar k_F$, $p_F \cong 1.6 \cdot 10^{-27}$ kg m/s taking $k_Fl \sim 1$ [10]. For the effective mass equal to the electron mass, one obtains $\xi_N \approx 6$ nm for junctions [7] with antiferromagnetic M interlayer with atomically thin F-layers. Following the analysis of the M/$S_d$ interfaces in Ref [33, 35], the amplitude of the superconducting current in S/$I_b$/M/$S_d$ structures between s- superconductor and mixed (s + d) superconductor is determined mostly by contribution of the s-components. Experimental data in Table 2 show that the MHS with the CSCO -interlayer have larger $I_cR_N$ than the MHS without it. Presence of s-wave component determines the critical current in the MHS, proportional to the product $\Delta_S\Delta_{Nb}$. However, the theoretical model does not describe resistive features of such MHS, namely, the exponential increase of the characteristic resistance $R_NA$ with $d_M$, and the deviation of the CPR from the sine-type. Again, resistance of MHS is determined by of the interface between the Au film and CSCO, which depends on the violation of stoichiometry of CSCO at the border and the relation of the Fermi velocities of these two materials [21]. The first does not depend on the thickness of the layer, while the second factor may cause an exponential change because of the gradient of conductivity over the thickness of the CSCO layer, as it is observed in the experiment. Deviation of the superconducting CPR from sinusoidal form due to appearance of second harmonic is a typical for Josephson structures with superconducting cuprate electrodes and is likely explained by their dominant d-wave superconducting order parameter [33, 36-38].

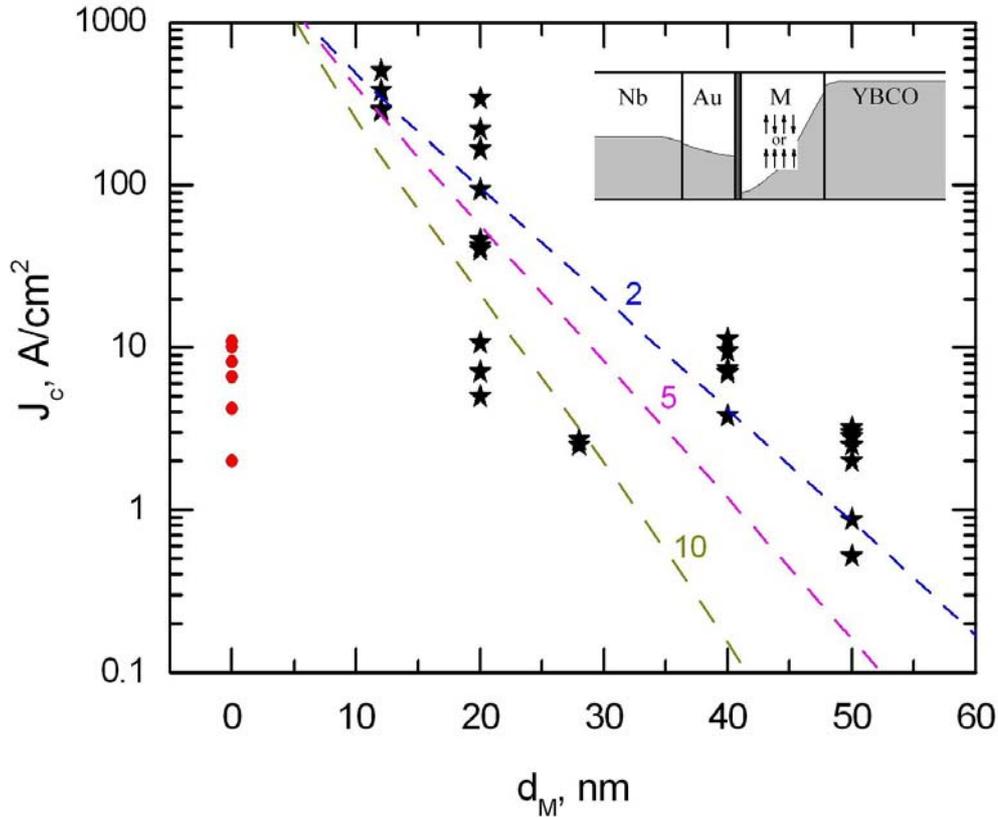

Fig.10. Experimental data (stars) and the calculation within MMS model (dashed lines) dependences of superconducting current density $j_c$ for AF-ordering in interlayer on the thickness $d_M$ for different level of exchange field $J_{ex}/\pi kT$. Coherence length of CSCO interlayer estimated from fitting experimental data with theoretical $j_c(d_M)$ dependence for $J_{ex}/\pi kT=2$ is $\xi_{AF}$ =7±1 nm. Filled circles show the data for MHS witout M-interlayer( $d_M$ ). The structure used for simulation is shown on inset. Variation of superconducting pair potential is shown schematically.

**5.3 Magnetic sensitivity**

Fig.11 presents the theoretical thickness dependences of the normalized critical current density $i_c(d_M)$ for the MHS with AF (solid lines) and F (dotted line) ordering for $J_{ex}/kT = 2, 3, 5$. Critical current is normalized by $I_{C0}$ - the critical current of hybrid heterostructure where M-interlayer is a normal metal. It is clearly seen that the $i_c$ for cases with F ordering are significantly smaller than for the AF-ordering. Consequently, if an external magnetic field disturbs the equilibrium collinear magnetization in the layers the critical current of MHS is reduced sharply. The lack of compensation for the magnetizations of layers due to the external field may cause suppression of the critical current as in the case of F-layer (see Fig. 11) There is also a qualitative change in the type of dependence $i_c(d_M)$, in particular, increasing the thickness $d_M$ is accomplished by transition between 0 - and $\pi$ - states.

According to G-K model [6] for S/AF/S structure the critical current $I_c$ depends on the canting of magnetization in layers induced, for example, by external magnetic field $H$ as following,

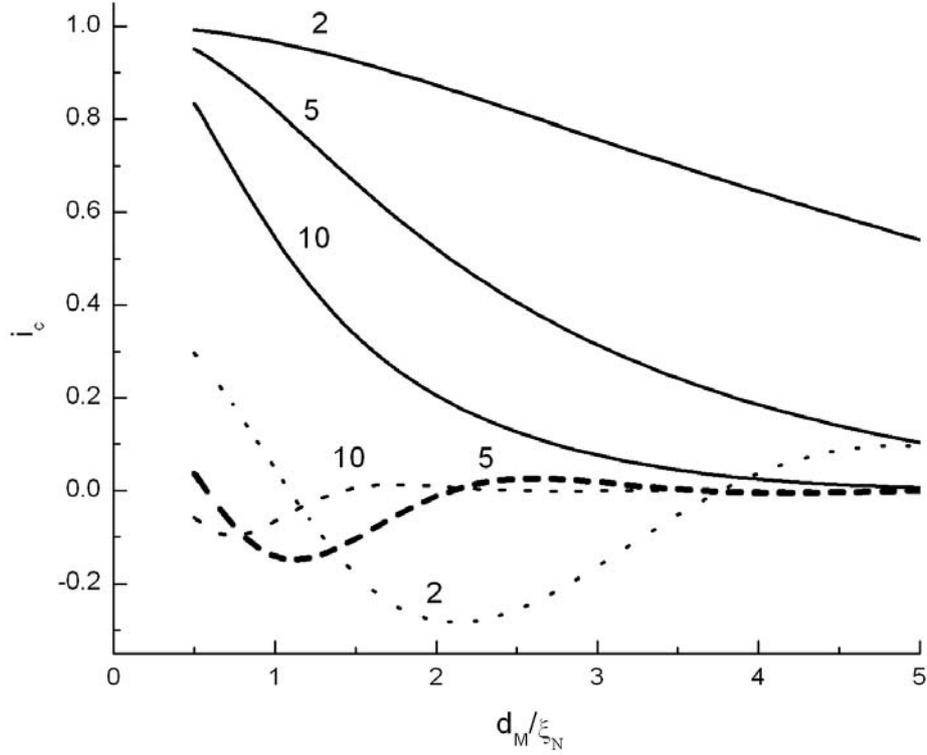

Fig.11. Theoretical thickness dependence of the normalized critical current density $i_c$ for for the AF (solid lines), and F (dotted line) orderings in interlayer for $J_{ex}/\pi kT = 2, 5, 10$. The data are normalized to the critical current density for the structure with interlayer consisting of a normal metal.

$$I_C \approx I_C^0 \left(\frac{2}{\pi\beta M_S}\right)^{1/2} \left|\cos(\beta M_S - \frac{\pi}{4})\right| \qquad (9)$$

The zeros of $I_C$ (9) correspond to relation $\beta M_S = \pi/4 + \pi n$ (n=1, 2..) and for $\beta = (t_0/T_c)(L/\xi_N) \gg 1$ ( $t_0$ is electron hopping parameters) the oscillations of $I_C$ could be observed at small canting (determinined by $M_S$ in (9)) [6]. Fig. 12 shows experimentally measured points of $I_C$ (H) for the MHS with the thickness of CSCO film interlayer $d_M$ = 50 nm. Suppose the canting induced by external nagnetic field the theoretical dependence $I_C$ (H) is calculated from eq.(9) normalized by $I_C^0$ (T = 4.2 K) taking the value of magnetic field $H_1$ at the first zero of experimental $I_C(H)$ as a fitting parameter.

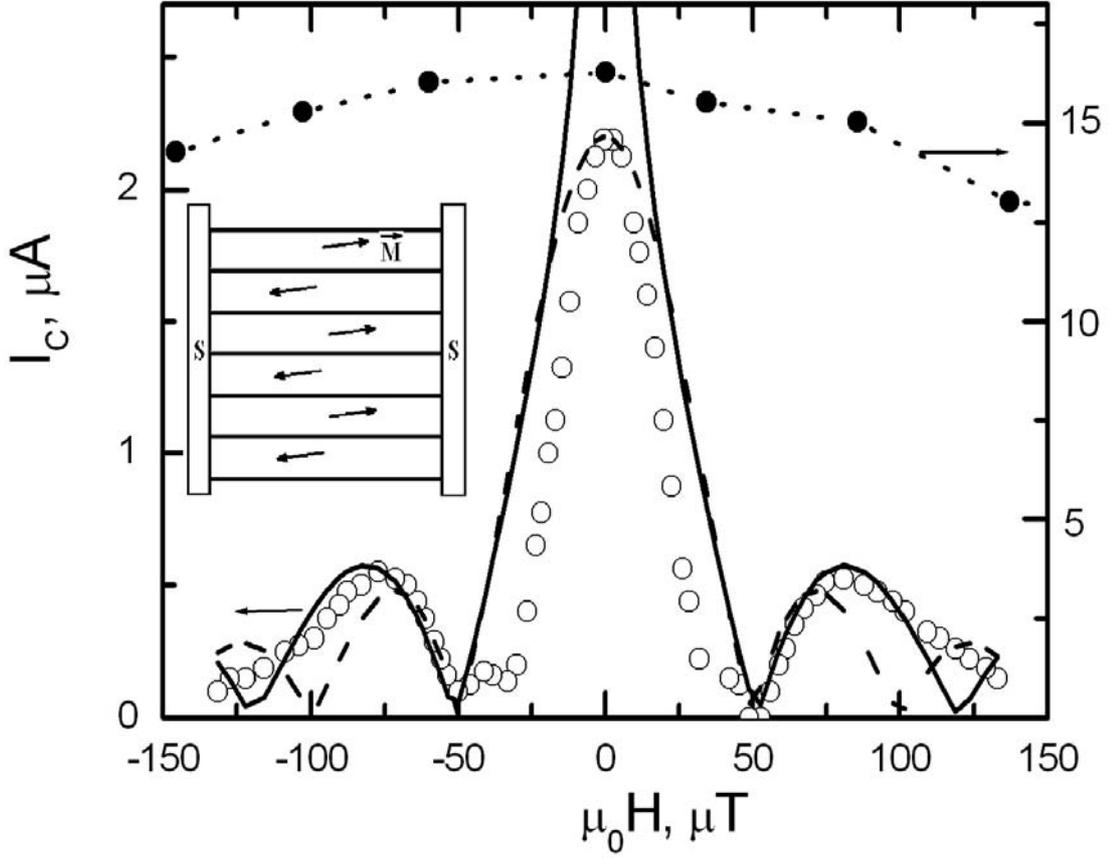

Fig.12. Magnetic-field dependence of the critical current $I_c$ (H) MHS with size L = 10 μm with CSCO-interlayer (open circles). Solid line is dependence obtained from formula (9), dotted line is Fraunhofer dependence (10). The calculated curves are normalized to the first zero of $I_c$ (H). Solid circles snow $I_c$(H) dependence for MHS without M-interlayer. Inset shows the sketch of Gorkov-Kresin model

It is seen that position of the second minimum of $I_C(H)$ differs significantly from the Fraunhofer pattern (see [39]):

$$I_C(H) = I_C^0 \left| \frac{\sin(\pi \Phi / \Phi_0)}{\pi \Phi / \Phi_0} \right|, \quad (10)$$

where $\Phi = \mu_0 H A_{ef}$ is magnetic flux through the MHS, $\mu_0$ is magnetic constant, $A_{ef} = L d_e$, $d_e = \lambda_{L1} + \lambda_{L2} + t$ is effective penetration depth of magnetic field, $\lambda_{Li}$ is London penetration depth for the electrodes of the MHS (i = 1, 2), t is barrier thickness. Zeros in (10) correspond to penetration of integer number of magnetic flux quanta $\Phi_0 = h/2e = 2.07 \cdot 10^{-15}$ Wb. The applicability of Eq.(9) at low H is limited, that results in observed deviation of the experimental points from solid line in Fig.12. Note, the absolute value of $H_1$ for the MHS with M-interlayer is significantly smaller than $H_1$ of the MHS without M-interlayer (see filled circles in Fig.12). A broadening of the secondary maxima of the $I_c(H)$ indicates that the observed period of $I_c(H)$ oscillations can be better fitted by G-K model than by Fraunhofer dependence (10). Such significant decrease in $H_1$ cannot be simply explained by a possible increase of the London penetration depth $\lambda_{L1}$ in YBCO due to a lower level of oxygen doping of the YBCO film near to the CSCO/YBCO interface seen from decreased critical temperature of YBCO down to $T_C \approx 80$ K. Note, results [45] show that less than 30% increase of $\lambda_L$ may happen if critical temperature of YBCO decreases to 40 K due to oxygen nonstoichiometry.

The $I_c(H)$ dependences are changed with increase of the of MHS size. At $L > 20$ μm: the critical current exhibits a maximum at low $H$, and amplitudes of the of $I_c(H)$ maxima are decreased in an oscillation manner with a period of about 1 μT. At $L = 50$ μm, the shape of the $I_c(H)$ seems corresponds to distributed Josephson junction model, although the distribution condition $L > 4\lambda_L$ is not met. The periodicity of this oscillating "fine structure" of $I_c(H)$ is well described by Eq. (9)[11].

### 6. Quasiparticle current in hybrid heterostructure

### 6.1. Hybrid heterostructure with AF/S$_d$ interface.

The voltage dependence of conductivity of MHS with CSCO interlayer at $T_c'<T<T_c$ may have two competing quasiparticle transfer channels like for MHS without M-interlayer [21]. Namely, a tunneling through the potential barrier at the Au/CSCO interface and a electron transport over Andreev bound states at the interface [34, 36]. The presence of Andreev bound states at the S'/M interface is confirmed by transformation of the σ(V) dependencies of the MHS, presented in Fig 13. At voltages V≥5 mV in temperature range $T_c > T > T'_c$ a conductance anomaly was observed for the tilted c-axis YBCO film related to low-energy Andreev bound states [21] (see the low curve on Fig.13). Earlier such anomaly was observed for MHS without CSCO layer when YBCO film had c-axis tilted orientation [21].

In general, the I-V curves of our Nb/Au/CSCO/YBCO MHS are of the hyperbolic shape at V≤ 1 mV, which is typical for Josephson junctions. No YBCO gap has been observed in the I-V curves. However, at $T < T_c'$ the dips on σ(V) at V≤ 2 mV are associated with the superconducting gap of Nb electrode (see Fig.13). Consequently, the d-wave component of the current is presented at the M/S$_d$ interface. At T <T'$_c$ for MHS on the *c*-oriented films no Andreev bound states were observed similar to the case of MHS without M-interlayer [21].

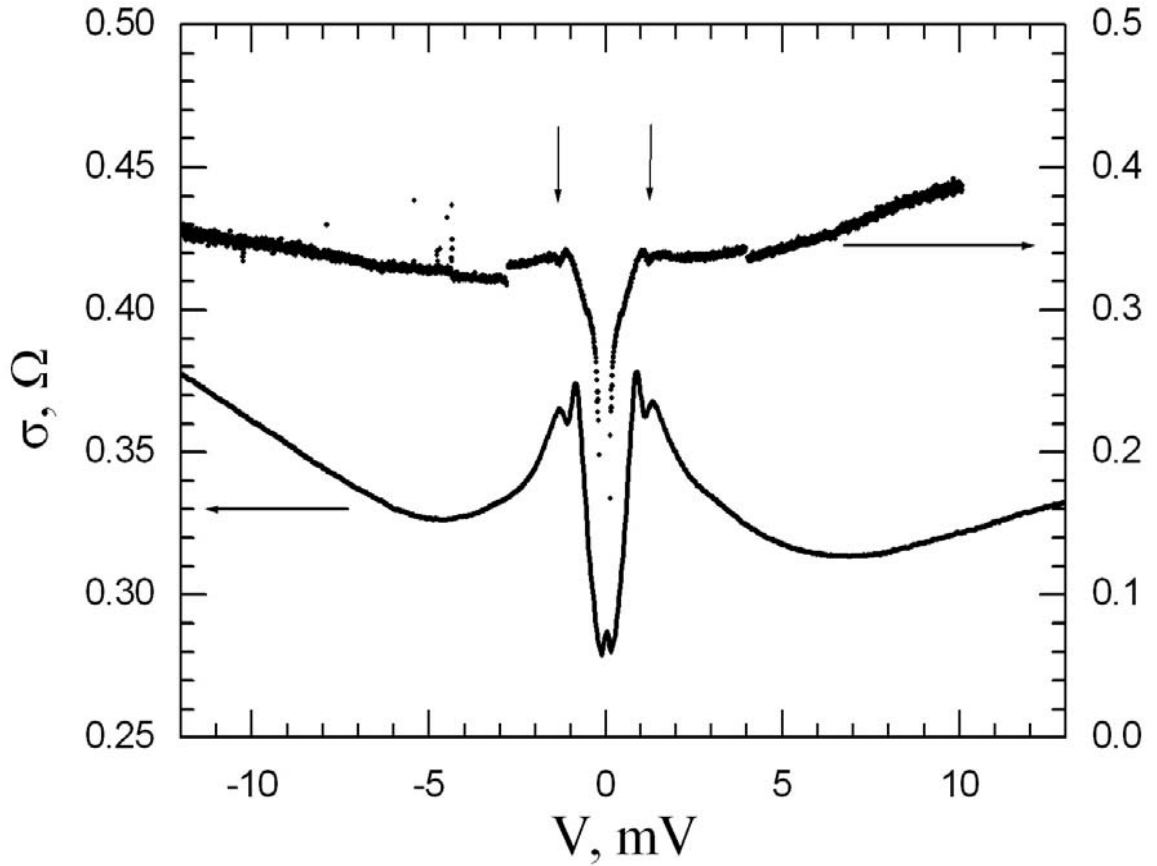

Fig.13. Voltage dependence of the electrical conductivity $\sigma(V)$ for Nb/Au/CSCO/YBCO MHS for (110)NGO (up curve) and tilted (below curve) substrates. Gap dips caused by the Nb electrode superconducting gap structure (marked by arrows) are observed.

### 6.2. Hybrid heterostructure with F/$S_d$ interface

For MHS with the M-interlayers made of manganite films no critical current was measured although the thickness of $d_M$ was reduced down to 5 nm. We used Ca and Sr doped manganite (LCMO and LSMO) as well as the underdoped LMO. Temperature dependences of resistance of MHS with manganite interlayers are presented in Fig.14. At high temperatures, R(T) is still determined mainly by the YBCO electrode. Contribution of manganite layer to the resistance of the MHS dominates at T ≤$T_c$ . The R(T) depends on the manganite doping level. Resistance of MHS with LMO interlayer increases monotonically with decrease the temperature (see Fig. 14), although such increase is slower than in the autonomous LMO films (see Fig.3).

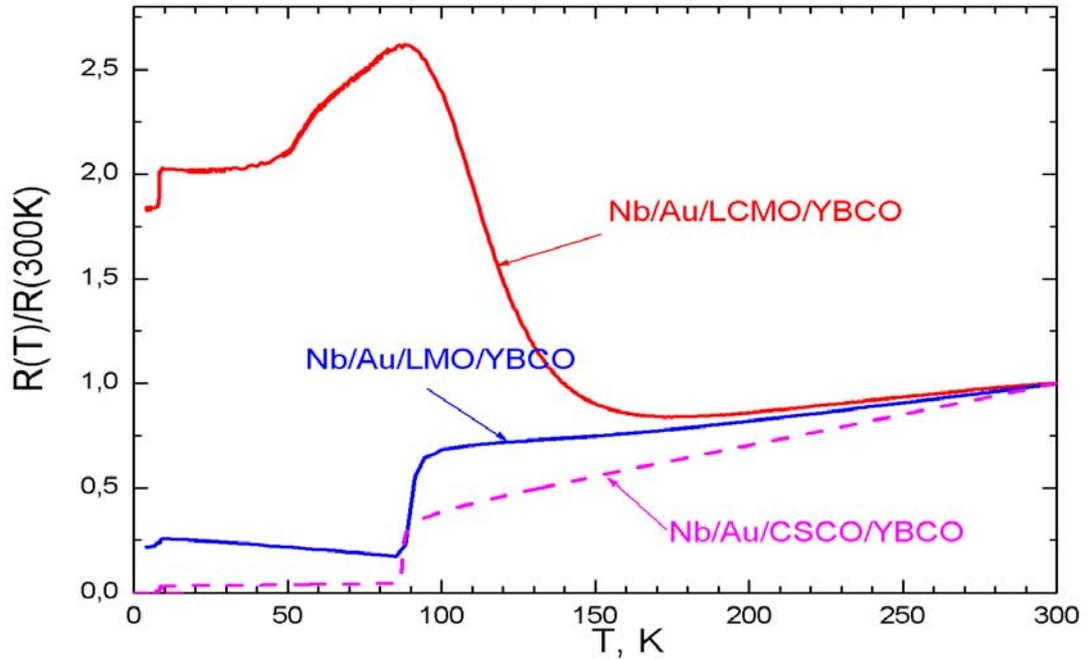

Fig.14. Temperature dependence of resistance *R(T)* for MHS with three types of M-interlayer: CSCO, LMO and LCMO. Strong deviation *R(T)* for LCMO interlayer due to metal-insulator transition is clearly seen.

The absolute value of MHS resistance with LMO interlayer is much smaller than the resistance of the autonomous LMO film calculated from the resistivity data (Fig.3). The similar reduction of resistivity of CSCO interlayer is discussed earlier (see part 4.1). Taking the estimated maximal resistance of Au/LSMO interface [46] and assuming that the contribution of M-layer resistance is small, we find that the determining factor in MHS resistance comes from LMO/YBCO interface. A similar situation occurs also for MHS with LCMO M-layer. However, the decrease in temperature of metal-insulator transition (accompanied by a transition from paramagnetic to ferromagnetic state) observed for MHS with small thickness of the M-layer [47] results in non-monotonous dependence of R(T) at T <$T_c$. (Fig.14). The resistive measurements show that the transparencies of LCMO(LMO)/ YBCO interfaces are small and do not differ much from the transparencies of Au/LCMO(LMO) interfaces. Thus, our Nb/Au/M/YBCO MHS with manganite M-interlayer are structure with two low transparent barriers which strongly suppress the critical current (proportional to the second degree of transparency) [48].

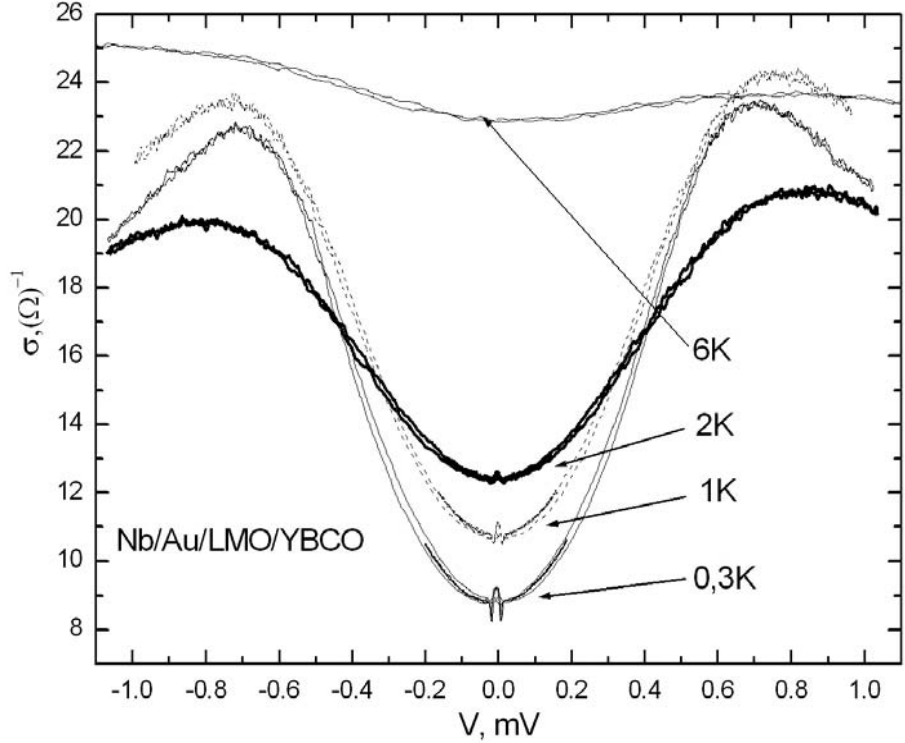

Fig.15. Voltage dependence of the electrical conductance $\sigma(V)$ of the Nb/Au/LMO/YBCO MHS for several temperatures. Gap features caused by the Nb electrode superconducting gap (marked by arrows) are clearly observed

The voltage dependences of conductivity $\sigma_d(V)$ of MHS with LCMO or LMO interlayers do not show the critical current down to low temperatures (0.3K) (Fig.15). This is probably happens due to the ferromagnetism of layers and the presence of the barrier at the $M/S_d$ interface. As shown in part 3.3, even in the case of LMO film, which in single crystals state usually shows the AF-ordering, its ferromagnetism is induced by oxygen nonstoichiometry. We observe increase of $\sigma_d$ caused by the superconducting gap of the Nb/Au bilayer $\Delta_{S'}$ at V≈1mV (Fig. 15)[6]. Consequently, superconducting correlations penetrate from Nb to Au, but are absent in the M-layer beeing highly supresessed. Such MHS could be considered as $SI_bF^*$ structures where the density of states in F*can be modified by the influence of $S_d$ electrode, but, however, it does not lead to the superconducting current in MHS. Indeed, we observed some pecularities on $\sigma_d(V)$ at low voltage V< 100 μV like published in [49,50] but we can not identify them by Andreev bound states.

---

[6] For qualitive estimation the position of gap dips the depairing factor should be taken into account

## 7. Conclusion

Experimental study of hybrid heterostructures with the interfaces of cuprate superconductor/cuprate antiferromagnet show the presence of anomalous proximity effect and very well manifested Josephson effect. The hybrid heterostructures with these interfaces show critical current density up to $10^3 A/cm^2$ at T=4.2 K and the characteristic voltage 100-200 μV when thickness of antiferromagnetic layer was 20-40 nm. The second harmonic of the current-phase relation of order 10-20% of the amplitude of the first harmonic one is observed. These samples show unusual for ordinary Josephsdon junctions high sensitivity to external magnetic field presumably caused by the canting of magnetization of individual layers under the influence of external magnetic field. The model describing the appearance of the anomalous proximity effect in the M-interlayer, consisting of magnetic layers with antiferromagnetic ordering, has been developed. Estimated the experimental and theoretical penetration depths of superconducting correlations were coincided within experimental error, and significantly exceed the values obtained in polycrystalline antiferromagnetic interlayer. At the same time, according to the calculations, the superconducting current decreases sharply with the appearance of ferromagnetic ordering in the interlayer, which is observed when ferromagnetic manganite were used as M-interlayer. The critical current in such structures was absent for structures with 5 nm thick layer at T= 0.3 K temperatures. However we do not exclude an influence of a sufficiently high barrier in the manganite/cuprate interface which limits the proximity effect.

The authors thank to T. Bauch, T. Claeson, I.M. Kotelyanski, K.E. Lakhmanski, F. Lombardi, V.A. Luzanov, A.M. Petrzhik and A. Pavolotsky for the discussions of the results and help in experiment, particular think to P. V. Komissinskiy for careful reading of manuscript. We gratefully acknowledge the partial support of this work by the Russian Academy of Sciences, Russian Foundation for Basic Research N08-02-00487, Scientific School Grant 5408.2008.2, International Scientific Technology Center project 3743.

References

1. A.I. Buzdin, Mod. Phys., **77**, 935 (2005).
2. F. S. Bergeret, A. F. Volkov, K. B. Efetov , Rev. Mod. Phys. **77**, 1321 – 1373 (2006).
3. L. N. Bulaevskii,., V. V. Kuzii, and A. A. Sobyanin, JETP Lett. **25**, 290 (1977)
4. V.V. Ryazanov, V.A. Oboznov, A.Yu. Rusanov et al., Phys. Rev. Lett. **86**, 2427 (2001)
5. J. Y. Gu *et al.*, Phys. Rev. Lett. **89**, 267001 (2002)
6. L.P. Gorkov, V.Z. Kresin, Appl. Phys. Lett **78**, 3657 (2001).
7. B. M. Andersen, Yu. S. Barash, S. Graser, P.J. Hirschfeld, Phys. Rev. **B77**, 054501 (2008); Phys.Rev. Lett., **96**, 117005 (2006).
8. A.V. Zaitsev, JEPT Lett, **90**, 521 (2009).
9. C. Bell, E.J. Tarte, G. Burnell, et al, Phys. Rev. **B68**, 144517 (2003).
10. P.V. Komissinskiy, G.A. Ovsyannikov, I.V. Borisenko, et al, Phys. Rev. Lett. **99**, 017004 (2007)
11. Y.V. Kislinskii, K.Y. Konstantinian, G.A. Ovsyannikov, et al, JEPT,. **106**, 800 ( 2008)
12. A.V. Zaitsev, G.A. Ovsyannikov, K.Y. Constantinian et al, (2010) (to be published)
13. A. Gozar, G. Logvenov, L. F. Kourkoutis, et al, Nature **455**, 782 (2008).
14. I. Bozovic, G. Logvenov, M.A.J. Verhoeven et al, Phys. Rev. Lett., **93**, 157002 (2004).
15. Y. Tarutani, T.Fukazawa, U.Kabasawa et al., Appl. Phys. Lett., **58**, 2707 (1991).
16. K.-U.Barholtz, M.Yu. Kupriyanov, U.Hubner et al, Physica C**334**, 175 (2000).
17. D.Vaknin, E.Caignol, P.K.Davis et al, Phys. Rev. B**39**, 9122 (1989).
18. M. Matsumura *et al.*, Phys. Rev. B**60**, 6285 (1999).
19. Yu.A. Izyumov, Yu.N. Scryabin, Phys. Usp, **44**, 109 (2001).
20. G. A. Ovsyannikov, I.V. Borisenko, P.V. Komissinski *et al.*, JETP Lett. **84**, 262 (2006).
21. P. V. Komissinskiy, G. A. Ovsyannikov, K. Y. Constantinian, et al. Phys. Rev B**78**, 024501 (2008).
22. G.A. Ovsyannikov, S.A. Denisuk, I.K. Bdikin et al., Physica C **408–410**, 616 (2004).
23. M. Varela, A.R. Lupini, S. Pennycook et al., Solid-State Electronics **47** 2245 (2003).
24. V. Peña, Z. Sefrioui, D. Arias, et al Phys. Rev. B **69**, 224502 (2004).
25. S. Stadler, Y.U. Idzerda, Z. Chen et al Applied Phys. Lett., **75**, 3384 (1999).
26. Y.Tokura, S.Koshihara, T.Arima et al. Phys. Rev.B**41**, 11657 (1990).
27. Q.Huang, A.Santoro, J.W. Lynn et al. Phys. Rev B **55**, 14987 (1997).
28. I.M. Fite, R. Szymczak, M. B. Baran et al. Phys. Rev B**68**, 014436 (2003).

29. G. A. Ovsyannikov, A. M. Petrzhik, I. V. Borisenko, et al, JEPT, **108**, 48 (2009).

30. S. Okamoto, A.J. Millis, Nature, **428**, 630 (2004).

31. J.C. Nie, P. Badica, M. Hirai et al Physica C **388–389**, 441 (2003).

32. S.J.L. Billinge, P.K. Davies, T. Egami, C.R.A. Catlow, Phys.Rev. B**43**, 10340 (1991).

33. P. V. Komissinski, E. Il'ichev, G.A. Ovsyannikov et. al. Europhys. Lett..**57**, 585 (2002).

34. A. Golubov, M. Yu. Kupriyanov, and E. Il'ichev, Rev. Mod. Phys. **76**, 411(2004).

35. K. Kuboki, Physica B**284-288**, 505 (2000).

36. T. Lofwander, V.S. ShumeikoV. S., G. Wendin, Superconducting Sci. Technol., **14** , R5 (2001).

37. E. Il´ichev, M. Grajcar, R. Hlubina et al. Phys. Rev. Lett., **86**, 5369 (2001).

38. M.H.S. Amin, A.N. Omelyanchouk, S.N. Rashneev et al. Physica B, **318**,162 (2002).

39. A. Barone, G. Paterno, N.-Y. Awaley-Interscience Publication John Wiley and Sons, 1982. 620 pages

40. A.F. Volkov, F.S. Bergeret, K.B. Efetov, Phys. Rev. Lett., **90**, 117006 (2006).

41. A.V. Zaitsev, JETP Lett, **83**, 277 (2006); *ibidem*, **88**, 521 (2008).

42. T. Yu. Karminskaya, M.Yu. Kupriyanov, and A.A. Golubov, JETP Lett., **87**, 570 (2008).

43. B. Crouzy, S.Tollis, D.A. Ivanov, Phys.Rev.B**76**, 134502 (2007).

44. D.A. Ivanov, Ya.V. Fominov, M.A. Skvortsov, P.M. Ostrovsky, arXiv, cond-mat:supr-con:0907.0113v (2009)

45. M.R. Trunin, Physics-Uspekhi, **48**, 979 (2005).

46. L. Mieville et al., Appl. Phys. Lett. **73**, 1736 (1998).

47. K.Dorr J.Phys.D:Appl.Phys. **39**, R125 (2006).

48. G.A. Ovsyannikov, G.E. Babayan, Physica B, **168**,239 (1991).

49. P. A. Kraus, A. Bhattacharya, and A. M. Goldman, Phys Rev B**64**, 220505(2001).

50. Z. Y. Chen, A. Biswas, I. Z utic´ et al Phys.Rev B**64**, 220505 (2001).